\documentclass[11pt]{article}
\usepackage{latexsym,bm,graphicx,color,xcolor,nicefrac,titletoc,enumerate,amsmath,amssymb,xfrac,xcolor,physics,cite,setspace}

\usepackage[nottoc]{tocbibind} 

\usepackage{mlmodern}
\usepackage[T1]{fontenc}

\usepackage{hyperref}
\hypersetup{linktocpage=true,colorlinks=true,linkcolor=blue,citecolor=blue,urlcolor=blue}

\usepackage[skip=3pt plus1pt, indent=20pt]{parskip}

\usepackage[bottom]{footmisc}

\usepackage{geometry}
\renewcommand{\thesection}{\arabic{section}}
\renewcommand{\thesubsection}{\arabic{section}.\arabic{subsection}}
\renewcommand{\thesubsubsection}{\arabic{section}.\arabic{subsection}.\arabic{subsubsection}}

\usepackage{titlesec}
\titleformat{\section}
{\normalfont\bfseries\sffamily}{\thesection}{0.3em}{}
\titleformat{\subsection}
{\normalfont\normalsize\bfseries}{\thesubsection}{0.3em}{}
\titleformat{\subsubsection}
{\normalfont\normalsize\bfseries}{\thesubsubsection}{0.3em}{}

\numberwithin{equation}{section}

\def\a{\alpha}
\def\b{\beta}
\def\g{\gamma}

\def\d{\delta}
\def\D{\Delta}
\def\e{\epsilon}

\def\f{\phi}

\def\P{\Psi}

\def\l{\lambda}

\def\m{\mu}
\def\n{\nu}
\def\r{\rho}
\def\s{\sigma}

\def\t{\tau}

\def\O{\Omega}

\def\pr{\prime}

\def\nn{\nonumber}
\def\qq{\quad\quad}

\newcommand{\sq}{\sqrt}
\newcommand{\sqdet}{\sq{-g}}

\newcommand{\cC}{\mathcal{C}}

\newcommand{\cG}{\mathcal{G}}
\newcommand{\cH}{\mathcal{H}}
\newcommand{\cL}{\mathcal{L}}

\newcommand{\cO}{\mathcal{O}}

\newcommand{\tT}{\widetilde{T}}

\newcommand{\onsh}{\eval_{\text{on-shell}}}

\newcommand{\IG}{I_{\text{G}}}
\newcommand{\IM}{I_{\text{M}}}
\newcommand{\BG}{B_{\text{G}}}
\newcommand{\BM}{B_{\text{M}}}

\newcommand{\fks}{\f_{\text{KS}}}

\newcommand{\mail}[1]{\href{mailto:#1}{{\tt #1}}}

\begin{document}
	
\begin{titlepage}       
	\vspace{5pt} \hfill 
		
		
	\begin{center}
		{\Large \bf \sffamily Classical double copy in the black hole mini-superspace}
	\end{center}
		
	\begin{center}
		\vspace{10pt}
			
		{{\bf \sffamily G{\"o}khan Alka\c{c},}${}^{a}\,${\bf \sffamily  Mehmet Kemal G\"{u}m\"{u}\c{s}}${}^{a}\,$ {\bf \sffamily and Mehmet Ali Olpak}${}^{b}$}
		\\[4mm]
			
		{\small 
		{\it ${}^a$Department of Physics Engineering, Faculty of Science and Letters,\\ Istanbul Technical University, Maslak 34469 Istanbul, Turkey}\\[2mm]
				
		
		{\it ${}^b$Department of Electrical and Electronics Engineering, Faculty of Engineering,\\ University of Turkish Aeronautical Association, 06790, Ankara, Turkey}\\[2mm]
				
		{\it e-mail:} {\mail{alkac@mail.com}, \mail{kemal.gumus@metu.edu.tr}, \mail{maolpak@thk.edu.tr}}
		}
		\vspace{2mm}
		\end{center}
		
		\centerline{{\bf \sffamily Abstract}}
		\vspace*{1mm}
		\noindent We give a novel formulation of classical double copy in the mini-superspace of static, spherically symmetric black holes where the map between the solutions of general relativity and Maxwell's theory can be realized in Boyer-Lindsquit coordinates. By employing the reduced action principle, we show that the double copy structure can be generalized to Lovelock gravities and quasi-topological gravities. However, the gravitational solutions are mapped to the purely electric solutions of Maxwell's theory with a difference: instead of a direct match between the Kerr-Schild scalar on the gravity side and the scalar potential on the gauge theory side, the scalar potential becomes a polynomial in the Kerr-Schild scalar, giving rise to a generalization of the Kerr-Schild double copy. We calculate the Regge-Teitelboim surface charges and prove that the mass of the black hole solution is identified with the electric charge corresponding to the Coulomb part of the gauge theory solution in the same way it does in the case of general relativity.
		 
		\vspace{3mm}
		\par\noindent\rule{\textwidth}{0.5pt}
		\tableofcontents
		\par\noindent\rule{\textwidth}{0.5pt}
		\newpage
		\pagestyle{empty}
\end{titlepage}

\section{Introduction}
The double copy idea emerged as relations between scattering amplitudes in gauge and gravity theories, where the quantum gravity amplitudes can be obtained by modifying the amplitudes in Yang-Mills theory \cite{Bern:2008qj}. A natural question is whether the notion of double copy can be extended to classical solutions. While such an extension is possible for a generic solution at a fixed order in perturbation theory \cite{Luna:2016hge}, certain algebraically special solutions of general relativity can be mapped to exact solutions of Maxwell's theory. This non-perturbative version of the double copy, known as the classical double copy, has two different formulations.

In the Kerr-Schild double copy \cite{Monteiro:2014cda}, a map to solutions of Maxwell's theory can be achieved thanks to the fact that, for metrics that can be written in the Kerr-Schild coordinates, the trace-reversed Einstein's equations with mixed indices are linear in the metric perturbation. This metric formulation can be naturally extended to spacetimes with a constant-curvature background \cite{CarrilloGonzalez:2017iyj}, and even with a generic curved background \cite{Alkac:2021bav}.

The second formulation, the Weyl double copy \cite{Luna:2018dpt, Godazgar:2020zbv}, is obtained by expressing the tensorial quantities characterizing the solutions in terms of spinors. In this spinorial formulation, the Weyl spinor of type D and type N solutions of general relativity in $d=4$ is related to a field strength spinor corresponding to a solution of Maxwell's theory defined on a suitable background spacetime. Recent findings in \cite{Easson:2021asd,Easson:2022zoh,Alkac:2023glx} suggest that it can match the known results in the Kerr-Schild double copy with some modifications to the original formula first proposed in \cite{Luna:2018dpt}.

In this work, motivated by the fact that the classical double copy is possible only for some algebraically special solutions, we give a third formulation by imposing the symmetries of the gravitational solution at the action level. To demonstrate the essentials of our novel formulation, we focus on one of the simplest, but also one of the most important, classes of solutions: static, spherically symmetric spacetimes. As we will see, our formulation is not only equivalent to the Kerr-Schild double copy but also allows an extension of the classical double copy to a class of higher-curvature gravity theories that form useful toy models for black hole physics in $d>4$, the quasi-topological gravities.

The outline of this paper is as follows: In section \ref{sec:KS}, we review the aspects of the Kerr-Schild double copy relevant to our work. In section \ref{sec:max}, we introduce the reduced action principle by applying it to Maxwell's theory. After a general discussion of the Regge-Teitelboim approach to conserved charges, we obtain the electric charge of the Coulomb solution in terms of an integration constant. Using the same ideas for general relativity in section \ref{sec:GR}, we present the mini-superspace approach to the classical double copy for static, spherically symmetric spacetimes and show its equivalence to the Kerr-Schild version. In section \ref{sec:QTG}, we show how the mini-superspace approach suggests a natural generalization of the Kerr-Schild double copy to quasi-topological gravities. Section \ref{sec:EGB} is devoted to an explicit realization of this generalization in 5d Einstein-Gauss-Bonnet theory. After this, the many-to-one nature of this generalization of the classical double copy is discussed and a comparison between our approach and previous works is given in Section \ref{sec:many}.  We end our paper with a summary of our results and some directions for future research in section \ref{sec:sum}. 

\section{Kerr-Schild double copy with a flat background metric}\label{sec:KS}
We consider stationary metrics that can be written in the following Kerr-Schild  form 
\begin{equation}\label{metKS}
g_{\m \n}=\eta_{\mu \nu}+\fks\, k_\mu k_\nu, \qq \qq \partial_0 g_{\mu \nu}=0,
\end{equation}
where the vector $k$ is null and geodesic with respect to both the flat background metric $\eta_{\m\n}$ and the full metric $g_{\m\n}$. Metrics that can be written in these coordinates have the important property that the Ricci tensor with mixed indices is linear in the deviation from the background such that the trace-reversed Einstein's equations are given by
\begin{equation}\label{reversed}
R^\m_{\ \n}=\frac{1}{2} \Big[\partial^\a \partial^\m\left(\fks\, k_\n k_\a\right)+\partial^\a \partial_\n\left(\fks\, k^\m k_\a\right)
 - \partial^2 \left(\fks\, k^\m k_\n\right)\Big]
=8 \pi G\, \tT^\m_{\ \n},
\end{equation}
where the tensor $\tT$ is related to the energy-momentum tensor as follows
\begin{equation}
\tT^\m_{\ \n}  = T^\m_{\ \n} -\frac{1}{d-2}\d^\m_{\ \n} T, \qq \qq T = T^\m_{\ \m}.
\end{equation}
The crucial observation for the Kerr-Schild double copy is that, when one chooses $k_0 = +1$, the $\m0$-components become Maxwell's equations \cite{Monteiro:2014cda}
\begin{equation}\label{maxwell}
\partial_{\nu}F^{\nu\mu}= \O_{D-2} \,J^{\mu},
\end{equation}
where $F_{\m\n}=2\partial_{[\m}A_{\n]}$ is the field strength corresponding to the gauge field
\begin{equation}\label{A}
A_\m = \fks\, k_\m,
\end{equation}
and the source is related to the energy-momentum tensor as
\begin{equation}\label{source}
J^\m = -\frac{16 \pi G}{\O_{D-2}}\, \tT^\m_{\ 0}.
\end{equation}
There exist various generalizations of this idea to cover more general classes of solutions of general relativity (see \cite{CarrilloGonzalez:2017iyj,Alkac:2021bav,Luna:2015paa} for examples). However; this most basic setup is sufficient for our main focus in this paper, which is static, spherically symmetric black hole solutions. In Boyer-Lindsquit coordinates, we can describe the metric of such a spacetime with the line element 
\begin{equation}\label{BL}
\dd{s}^2 = - f(r) \dd{t}^2+\frac{1}{f(r)} \dd{r^2}+ r^2 \dd{\O}^2_{D-2},
\end{equation}
where $\dd{\O}^2_{D-2}$ is the metric of a $(D-2)$-dimensional unit sphere. While more general solutions with $g_{tt} g_{rr} \neq -1$ are possible for some matter couplings and, also as vacuum solutions in some modified gravity theories, we contend ourselves with this simple form since it is enough to discuss all the essential features of the mini-superspace approach. After the following coordinate transformation
\begin{equation}
\dd{t} \to \dd{t} - \frac{1-f(r)}{f(r)} \dd{r},
\end{equation}
the metric can be written in the Kerr-Schild coordinates where the line element for the flat background spacetime with metric $\eta_{\m\n}$, the $k$ vector, and the Kerr-Schild scalar $\fks$ are given by
\begin{align}
\dd{\bar{s}}^2 &= \eta^{\mu \nu} \dd{x}^\m \dd{x}^\n= -\dd{t}^2+\dd{r}^2 + r^2 \dd{\O}^2_{D-2},\label{back}\\
k &= k_\m \dd{x}^\m = \dd{t}+\dd{r},\label{k}\\
\fks &= 1- f(r)\label{phitof}.  
\end{align}
As a result, any solution of general relativity with a line element of the form \eqref{BL} can be mapped to a solution of Maxwell's theory where the corresponding gauge field and source are given in (\ref{A}, \ref{source}) respectively. For a static, spherically symmetric solution whose Kerr-Schild scalar $\fks$ is related to the metric function as in \eqref{phitof}, one obtains a purely electric solution in Maxwell's theory defined on the Minkowski spacetime in the spherical coordinates \eqref{back}, where the scalar potential $\f$ is identified with the Kerr-Schild scalar $\fks$ of the gravitational solution. The simplest example of the map is provided by the vacuum solution of general relativity, the Schwarzschild black hole solution whose metric function is given by
\begin{equation}
f(r) = 1 - \frac{16 \pi G  M}{(D-2) \O_{D-2}r^{D-3}},
\end{equation}
where $M$ is the mass of the black hole. In $D\geq 4$, it is mapped to the Coulomb solution of Maxwell's theory with the gauge field\footnote{In 3d, as very well-known, a negative cosmological constant is required to obtain a black hole solution \cite{Banados:1992wn}. A gravitational solution corresponding to the Coulomb solution can only be obtained by matter coupling \cite{CarrilloGonzalez:2019gof,Gumus:2020hbb,Alkac:2021seh,Alkac:2022tvc}.}
\begin{equation}\label{AKS}
A= A_{\m} \dd{x}^\m = \frac{Q}{(D-3)r^{D-3}}\left(\dd{t}+\dd{r}\right),
\end{equation}
where the electric charge $Q$ of the point particle in Maxwell's theory is related to the black hole mass as
\begin{equation}\label{QtoM}
Q = \frac{(D-3)16 \pi G  M}{(D-2)\O_{D-2} }.
\end{equation}

In the next two sections, starting from the vacuum case, we show how one can obtain this map in the mini-superspace approach without any need to consider the gravitational solution in the Kerr-Schild coordinates.

\section{Mini-superspace approach to Maxwell's theory}\label{sec:max}
Although gauge fixing at the action might be in general problematic, according to a theorem due to Palais, one can still obtain the solution with the desired symmetries in some cases \cite{Palais:1979rca} (see also \cite{Fels:2001rv,Frausto:2024egp}). As a simple exercise, let us take  free Maxwell's theory described by the action
\begin{equation}\label{actmax}
\IM= - \frac{1}{4 \Omega_{D-2}} \int \dd^D x \sqrt{-\eta}\, F_{\m\n} F^{\m\n},
\end{equation}
defined on the Minkowski spacetime in spherical coordinates with the line element \eqref{back}, and study the solutions of the form
\begin{equation}\label{ABL}
A = A_\m \dd{x}^\m = \f(r) \dd{t},
\end{equation}
where $\f$ is the scalar potential. Checking the $0$-th component of the covariant form of the field equations, one of course finds the free Poisson's equation, i.e., $\partial_{\n} F^{\n 0} = - \vec{\nabla}^2 \phi = 0$ whose solution is the Coulomb potential $\phi = A + \frac{B}{r}$ ($A,B$ : constant).

Instead of working with the covariant form of the field equations, we can try to obtain a reduced action for the mini-superspace formed by the solutions of the form \eqref{ABL}. Inserting the ansatz \eqref{ABL} directly into the action \eqref{actmax}, one finds
\begin{equation}\label{maxredraw}
	\IM =  \frac{\D t}{2} \int \dd{r} r^{D-2} {\phi^\pr}^2,
\end{equation}
where we have used $\int_{t_1}^{t_2} \dd{t} = t_2 - t_1 \equiv \D t$ and the integration over the angular variables leads to a $\Omega_{D-2}$ factor, which cancels the pre-factor in the action \eqref{actmax}. Since we have assumed only a radial dependence for the scalar potential, we obtained a one-dimensional minisuperspace model for the type of solution that we want to obtain. Varying the reduced action with respect to $\phi$, we obtain its equation of motion as
\begin{equation}\label{phieqn2}
	\dv{r}\left(r^{D-2} \f^\prime\right) = 0,
\end{equation}
which is the free Poisson's equation for a scalar potential depending only on the radial coordinate $r$. Although it might seem rather trivial at this stage, the fact that we are able to obtain the correct equation of motion from the reduced action also implies that \emph{all} the physical properties of the solution can also be derived from the reduced action. 

According to the Hamiltonian Regge-Teitelboim method \cite{Regge:1974zd}, the conserved charges of a gauge theory are derived from the surface term that needs to be added to the Hamiltonian generator to ensure well-defined functional derivatives. This is required for the gauge symmetries of the theory to be properly generated through Poisson brackets. 

The easiest way to find the required surface term is to study the Hamiltonian form of the action. For a general gauge theory, it takes the following form
\begin{equation}\label{actgen}
	I = \int \dd^D x \left(p_i \dot{q}^i - \cH_0 + \l^a \Phi_a\right) + B,
\end{equation}
where $\cH_0$ is the dynamical part of the Hamiltonian density and $\Phi_a = 0$ ($a = 1, \cdots, N$) correspond to $N$ constraints arising from the gauge invariance and they are imposed via the Lagrange multipliers $\l^a$. $B$ is a boundary term whose role will be explained soon. This form is true for the action of a relativistic point particle, Maxwell's theory, Yang-Mills theory and general relativity. The reader is referred to the excellent review of Banados and Reyes \cite{Banados:2016zim} for details. The canonical generator is defined as 
\begin{equation}
	H = \int \dd^{D-1} x \left(\cH_0 -  \l^a \Phi_a \right) + \cC,
\end{equation}
where $\cC$ corresponds to the required surface term. If we choose the boundary term $B$  in \eqref{actgen} such that the variation of the action \eqref{actgen} vanishes on-shell, i.e. $\d I\onsh = 0$, this guarantees that the Hamiltonian has well-defined functional derivatives. 

The crucial observation of Regge and Teitelboim is that when the surface term $\cC$ is evaluated for a solution at spatial infinity, it yields an expression for the conserved charges corresponding to the gauge symmetries of the theory, which schematically takes  the following form
\begin{equation}
	\cC = \l^a_\infty C_a.
\end{equation}
Here, $ \l^a_\infty$'s are the Lagrange multipliers evaluated at infinity and $C_a$'s are the conserved charges. For all the gauge theories that we have mentioned, they always appear in such conjugate pairs. Since we have an additional time integral in the action \eqref{actgen}, this means the boundary term is related to the conserved charges as
\begin{equation}\label{chargedef}
\eval{\d B}_{\infty} = - \D t \l^a_\infty  \d C_a.
\end{equation}
The conserved charges are naturally related to the integration constants that arise when solving the field equations. By taking variations of the fields with respect to these integration constants and using the identification \eqref{chargedef}, one can easily find the relation between the integration constants and the conserved charges.

After this general discussion, we can now investigate the Hamiltonian form of the action for Maxwell's theory given in \eqref{actmax}, and understand the simplification offered by the reduced action. Considering a general field configuration $A^\m = (\f, A^i)$ and writing the Minkowski spacetime in Cartesian coordinates, it reads
\begin{equation}\label{actmaxham}
	\IM =\int \dd^D x\left[\pi_i \dot{A}^i-\left(\frac{1}{2} \pi_i \pi^i+\frac{1}{4} F_{i j} F^{i j}\right)+\phi\,\cG \right] + \BM.
\end{equation}
Here, canonical momenta $\pi_i$ are just the components of the electric field since $
\pi_i \equiv \frac{\partial \mathcal{L}}{\partial \dot{A}_i}=\dot{A}_i-\partial_i A_0
$. $\cG = \partial_i \pi^i$ is the Gauss' law constraint of electromagnetism and the scalar potential $\f$ is the corresponding Lagrange multiplier. One can proceed with this form of the Hamiltonian action to derive a general expression for the surface term $\cC$ (and the conserved charge). However, for a more complicated theory such as general relativity and its extensions that we consider in this paper, it is much more practical to do it for the solution under consideration only and this is possible provided that the solution can be obtained from a reduced action consistently.

Going back to our reduced action in \eqref{maxredraw}, we immediately see that it is not in the Hamiltonian form yet. In the one-dimensional mini-superspace, we need to introduce a momentum such that the variation with respect to the Lagrange multiplier $\f$ will lead to a constraint on the momentum. Also, the remaining terms should contain the square of the momentum as we have in the dynamical Hamiltonian. This way, we have the same structure as we have in the general Hamiltonian action \eqref{actmaxham}. 

We can make the following ansatz for the reduced Hamiltonian form of action
\begin{equation}
\IM=  \D t \int \dd{r} \left(\f\,  p^\prime +  \chi\, p^2 \right),
\end{equation}	
where we introduced a rescaled canonical momentum $p$ and the function $\chi$ can be used to make sure that it gives the reduced action in \eqref{maxredraw} (up to boundary terms) when the definition of the momentum $p$ is inserted back into this action. After some simple calculations, one finds $\chi = -\frac{1}{2 r^{D-2}}$ and we arrive at the following action
\begin{equation}
\IM=  \D t \int \dd{r} \left(\f\,  p^\prime - \frac{p^2}{2 r^{D-2}} \right) +\BM.
\end{equation}
The equations of motion and their solutions are as follows:
\begin{align}
p^\prime =0, \qquad &\Rightarrow \qquad p=p_0,\label{peqn} \\
\f^\prime = - \frac{p}{r^{D-2}},\qquad &\Rightarrow \qquad \f = \f_{\infty}  +  \frac{p_0}{(D-3)r^{D-3}}.\label{phieqn}
\end{align}
Note that our Gauss' law constraint imposed by the Lagrange multiplier $\f$ now takes the form $\cG = p^\pr = 0$.

With these at hand, we can now calculate the conserved electric charge $Q$ in terms of the integration constant $p_0$. In this case, the Lagrange multiplier $\f$ evaluated at infinity $\f_\infty$ and the integration constant $p_0$ form a conjugate pair, and the boundary term
\begin{equation}
\d \BM = - \D t \f \d p,
\end{equation}
which is obtained from the condition $\d \BM|_{\text{on-shell}} = 0$, is identified with
\begin{equation}
\eval{\d \BM}_{\infty} = - \D t \f_{\infty} \d Q,
\end{equation}
where $Q$ is the electric charge. Using $\d p|_\infty =\d p_0$, which follows from the solution for the momentum \eqref{peqn}, one finds the electric charge as
\begin{equation}
Q = p_0.
\end{equation}
This just tells us the scalar potential in terms of the electric charge is $\f = \f_{\infty}  +  \frac{Q}{(D-3)r^{D-3}}$, where $\f_{\infty}$ can now be set to zero and the electric field takes the Coulomb form $E = \frac{Q}{r^{D-2}}$. One might think this is too much work to obtain such a simple result. However, for general relativity and some of its generalizations, it offers quite a big simplification in obtaining the conserved charges, especially for spherically symmetric solutions. In the next section, we will see that there is no need for defining a momentum in such cases and the reduced action that is obtained by inserting the ansatz for the solution into the action can be directly used after integrating by parts a few times.

\section{Mini-superspace approach to general relativity}\label{sec:GR}
In order to obtain the Hamiltonian form of the action for general relativity, we use the Arnowit-Deser-Misner (ADM) decomposition \cite{Arnowitt:1962hi} of the line element given by
\begin{equation}
	\dd{s}^2 = -N_\perp^2 \dd{t}^2 + g_{i j} (\dd{x^i} + N^i \dd{t}) (\dd{x^j} + N^j \dd{t}),
\end{equation}
where $N_\perp$ is the lapse function, $N^i$ is the shift vector and $g_{ij}$ is the $(D-1)$-dimensional metric on $t = \text{constant}$ surfaces. ADM showed that the Einstein-Hilbert action, which is given by
\begin{equation}\label{actEHcov}
I_{\text{EH}} = \frac{1}{16 \pi G} \int \dd^D{x} \sqdet\, R,
\end{equation}
takes the following Hamiltonian form
\begin{equation}\label{actgrham}
\IG =\int \dd^D x \left[\Pi_{ij}  \dot{g}^{ij} - N_\perp \cH - N^i \cH_i\right]+ \BG,
\end{equation}
which is in agreement with the general form given in \eqref{actgen}. As a generic property of diffeomorphism invariant theories, there is no dynamical part of the Hamiltonian ($\cH_0$). The lapse function $N_\perp$ and the shift vector $N^i$ are the Lagrange multipliers imposing the Hamiltonian constraint $\cH = 0$ and the diffeomorphism constraint $\cH_i = 0$. More details on the general construction can be found in the original work or in the review \cite{Banados:2016zim}.

Instead of a general analysis, following \cite{Crisostomo:2000bb}, we
now show how to obtain the  static, spherically symmetric spacetimes with the line element \eqref{BL} from a one-dimensional mini-superspace
model and find the relation between the integration constant and the conserved charge via the Regge-Teitelboim approach \cite{Regge:1974zd}. In order to derive the reduced action for such spacetimes, we insert the metric with the line element
\begin{equation}\label{BLwithN}
\dd{s}^2 = - N^2(r) \left(1-\fks\right) \dd{t}^2+\frac{1}{\left(1-\fks\right)} \dd{r^2}+ r^2 \dd{\O}^2_{D-2},
\end{equation}
into the Einstein-Hilbert action \eqref{actEHcov}. Note that we work in the Boyer-Lindsquit coordinates but write the metric function $f(r)$ in terms of the Kerr-Schild scalar $\fks$, which we derived in \eqref{phitof}. After integrating by parts, one obtains the reduced action of our gravitational theory as
\begin{equation}\label{redGR}
\IG = \frac{(D-2) \Omega_{D-2} \D t}{16 \pi G} \int \dd{r} N \Psi^\prime + \BG,
\end{equation} 
where 
\begin{equation}\label{psiGR}
\P = r^{D-3}\fks,
\end{equation}
and we again introduce a boundary term $\BG$ such that one has a well-posed variational problem, i.e., $\d\IG|_{\text{on-shell}} = 0$. It is important to emphasize that the function $N(r)$ should be introduced in the line element \eqref{BLwithN}, otherwise one would only get a boundary term in the reduced action, and as a result, cannot deduce the solution of the form \eqref{BL}. This procedure for obtaining such solutions from the reduced action is valid for a wide range of gravity theories and we refer the reader to \cite{Deser:2003up} for various examples. The equations of motion that follow from the reduced action \eqref{redGR} and their solutions are given by
\begin{align}
N^\prime = 0, \qquad &\Rightarrow \qquad  N=N_\infty,\label{Neqn} \\
\Psi^\prime=0, \qquad  &\Rightarrow \qquad  \Psi = m.\label{psieqn}
\end{align}
As we see, the Lagrange multiplier $N$ leads to the Hamiltonian constraint $\cH \sim \Psi^\pr = 0$ and our reduced action \eqref{redGR} is already in the Hamiltonian form [see the general expression in \eqref{actgrham}]. For the type of solutions described by the line element \eqref{BLwithN}, there is no need to introduce a momentum as we did for Maxwell's theory and we can directly proceed with this form of the reduced action.

From the function $\P$ given in \eqref{psiGR}, the equation satisfied by the Kerr-Schild scalar $\fks$ and its solution follow as
\begin{equation}\label{fkssol}
\dv{r} \left(r^{D-3} \fks \right) = 0, \qquad \Rightarrow \qquad \fks = \frac{m}{r^{D-3}},
\end{equation}
where $m$ is an integration constant related to the mass $M$ of the black hole and, it forms a conjugate pair with the constant $N_\infty$ in the Regge-Teitelboim approach. From the condition $\d\IG|_{\text{on-shell}} = 0$, we find that the variation of the boundary term should be given by
\begin{equation}\label{delBG}
\d \BG =  - \frac{(D-2) \D t\, \Omega_{D-2} N_\infty }{16 \pi G}  \d \P,
\end{equation}
which, at infinity, should match with
\begin{equation}\label{massdef}
\eval{\d \BG}_{\infty} = - \D t N_\infty \d M,
\end{equation}
where $M$ is the black hole mass. The variation of the function $\P$ at infinity can be obtained from the equation of motion \eqref{psieqn} as 
\begin{equation}\label{delpsiinf}
\eval{\d \Psi}_\infty = \d m.
\end{equation}
Inserting this in \eqref{delBG} and comparing the result with \eqref{massdef}, one obtains  the mass of the black hole solution $M$   in terms of the integration constant $m$  as
\begin{equation}\label{M}
M = \frac{(D-2) \Omega_{D-2}\, m}{16 \pi G }.
\end{equation}

Note that the functional forms of the solutions for the Kerr-Schild scalar \eqref{fkssol} and the scalar potential \eqref{phieqn} are the same when the value of the scalar potential at infinity is set to zero, i.e., $\f_\infty=0$. The fact that this is not a coincidence can be seen as follows: Denoting the equations of motion for the Kerr-Schild scalar $\fks$ \eqref{fkssol} and the scalar potential \eqref{phieqn2}  $\text{EOM}(\fks)$ and $\text{EOM}(\f)$ respectively, one finds the following relation between the two
\begin{equation}\label{fromgrtom}
\left[r\, \dv{r}-(D-4)\right] \text{EOM}(\fks) = \text{EOM}(\f).
\end{equation}
The additional integration constant $\f_\infty$ appears in the solution for the scalar potential $\f$ in \eqref{phieqn} because one has a second-order differential equation after the differentiation of the equation for the Kerr-Schild scalar $\Phi(\fks)$. Therefore, the two equations are not independent and must have the same solution when $\f_\infty=0$. In conclusion, we have
\begin{equation}
\f = \fks, \quad \text{when } \f_\infty = 0,
\end{equation}
which also implies the following relation between the integration constants
\begin{equation}\label{intcons}
m = \frac{p_0}{D-3}.
\end{equation}
Together with the relation between the integrations constants \eqref{intcons}, this implies the relation \eqref{QtoM} between the black hole mass $M$ and the electric charge $Q$ of the point particle sourcing the Coulomb solution that we obtained in the Kerr-Schild double copy.

With this, we have completed the derivation of the map between the Schwarzschild solution and the Coulomb solution together with the relation between the conserved charges. Incorporation of the static, spherically symmetric sources is pretty straightforward. On the gravity side, one can easily show that the equation for the Kerr-Schild scalar obtained from the reduced action \eqref{fkssol} is related to the components of the Einstein tensor as follows:
\begin{equation}\label{redeqn}
\dv{r} \left(r^{D-3}  \fks\right) = -\frac{2}{D-2}\, r^{D-2}G^t_{\ t} = -\frac{2}{D-2}\,  r^{D-2} G^r_{\ r}.
\end{equation}
Therefore, defining
\begin{equation}\label{fkssourced}
\e(r)\equiv  T^t_{\ t}=  T^r_{\ r},
\end{equation}
one would have 
\begin{equation}\label{fkssourced}
\dv{r} \left(r^{D-3}\fks\right)=  -\frac{16 \pi G}{D-2}\, r^{D-2}\e .
\end{equation}

For Maxwell's theory, one can use the sourced version of the action \eqref{actmax}
\begin{equation}
\IM=  \int \dd^D x \sqrt{-\eta}\, \left[- \frac{1}{4 \Omega_{D-2}} F_{\m\n} F^{\m\n} + A_\m J^\m \right],
\end{equation}
which leads to the covariant form of Maxwell's equation given in \eqref{maxwell},
and take a source of the form
\begin{equation}
J = J^\m \partial_\m = \r(r) \partial_t,
\end{equation}
where $\r(r)$ is the charge density. The equation for the scalar potential becomes
\begin{equation}\label{fsourced}
\dv{r}\left(r^{D-2} \f^\prime\right) = - \O_{D-2}\, r^{D-2} \r,	
\end{equation}
which is, of course, just the sourced Poisson's equation. Knowing how the left-hand-sides of the equations \eqref{fkssourced} and \eqref{fsourced} are related from \eqref{fromgrtom}, it is straightforward to deduce the following relation between the sources 
\begin{equation}
\r = \frac{16 \pi G\left(r \e^\prime + 2 \e\right)}{(D-2)\O_{D-2}}.
\end{equation}
With the matter coupling, both the Kerr-Schild scalar $\fks$ and the corresponding scalar potential $\f$ are modified. However, the relation between the black hole mass $M$ and the electric charge $Q$ corresponding to the Coulomb part of the gauge theory solution \eqref{QtoM} remains intact.

This finalizes our alternative formulation of the classical double copy for static, spherically symmetric solutions of general relativity. We would like to emphasize that the calculation in the gravity side was performed in the Boyer-Lindsquit coordinates and the corresponding gauge field \eqref{ABL} take a different form than the one in the Kerr-Schild double copy \eqref{AKS}. Since the $r$-component of the latter can be gauged away, physically they describe the same solution, which can be seen by calculating the electric field, which in both cases gives $e = - F_{rt} = -\f^\prime.$
\section{Generalization to quasi-topological gravities}\label{sec:QTG}
Note that, in establishing the map between solutions, the special form of the reduced action \eqref{redGR} played a crucial role, thanks to which, the Kerr-Schild scalar obeys a first-order differential equation. In the covariant form of the field equations, described by the Einstein tensor, while some components are algebraically related to the one following from the reduced action that is given in \eqref{redeqn}, others are second-order differential equations which are solved by the solution obtained from the reduced action. Therefore, the particular form of the reduced action is directly related to the fact that general relativity has second-order field equations for the class of metrics described by the line element \eqref{BL}. This is, in some sense, trivial since general relativity has second-order field equations for a generic metric, which is described by the Einstein tensor. It is a characteristic property of the well-known Lovelock gravities \cite{Lanczos:1938sf,Lovelock:1971yv,Lovelock:1972vz} (see \cite{Padmanabhan:2013xyr} for a review of important properties) described by the Lagrangians
\begin{equation}\label{Lm}
\mathcal{L}_m=\frac{1}{2^m}\delta^{\mu_1\nu_1\cdots\mu_m\nu_m}_{\rho_1\sigma_1\cdots\rho_m\sigma_m}R^{\rho_1\sigma_1}_{\ \ \ \ \  \mu_1\nu_1}\cdots R^{\rho_m\sigma_m}_{\ \ \ \ \ \ \ \mu_m\nu_m},
\end{equation}
which have a non-trivial contribution to the field equations in $D>2m$, leaving the general relativity as the unique theory with second-order field equations for a generic metric in $D=4$. Motivated by this special form of the reduced action, which is also shared by Lovelock gravities with a modified $\P$ function, one can ask for which theories the reduced action computed for the line element \eqref{BLwithN} have the same form such that the field equations are second-order differential equations for metrics of the form \eqref{BL}. Answering this question results in the so-called quasi-topological gravities which have second-order field equations for such spacetimes, but not for a generic spacetime, and naturally include Lovelock gravities as a sub-class. These theories share the property of Lovelock gravities in  that they admit a unitary, massless spin-2 excitation around any of their constant-curvature vacua. There is an extensive literature on the subject and we refer the reader to \cite{Oliva:2010eb,Myers:2010ru,Dehghani:2011vu,Oliva:2011xu,Ahmed:2017jod,Cisterna:2017umf,Bueno:2019ycr,Bueno:2019ltp,Bueno:2022res,Moreno:2023rfl,Bueno:2024dgm} for details. The reader who is only interested in the form of the Lagrangians can check the supplemental material of \cite{Bueno:2024dgm}. What is important for our discussion is that, in each $D\geq5$, unlike Lovelock Lagrangians that truncate after the order $m = \left[\frac{D}{2}\right]$, there are infinitely many quasi-topological Lagrangians that possess a reduced action in the form \eqref{redGR} where the function $\P$ can be written as
\begin{equation}\label{psigen}
\Psi = r^{D-3} \psi, \qq  \psi = \sum_{n=1} ^{n_{\text{max}}} c_n \frac{\fks^n}{r^{2(n-1)}},
\end{equation}
with $c_1=1$, corresponding to the contribution from the Einstein-Hilbert action, and $c_{n>1}$'s are related to the coupling constants of the higher-order quasi-topological Lagrangians. Having the same form of the reduced action, the equations of motion for the functions $N$ and $\P$ remain the same, as given in (\ref{Neqn}, \ref{psieqn}). With a modified $\P$ function, one now has the following equation for the Kerr-Schild scalar
\begin{equation}\label{fksgen}
\dv{r} \left( r^{D-3} \psi \right) = 0 \quad \Rightarrow \quad \psi =  \sum_{n=1} ^{n_{\text{max}}} c_n \frac{\fks^n}{r^{2(n-1)}} = \frac{m}{r^{D-3}},
\end{equation}
where $m$ is again an integration constant related to the black hole mass $M$ and forms a conjugate pair with the integration constant $N_\infty$. Recall that, in our conserved charge analysis, we have used the variation of the function $\P$ at infinity, which we found from the equation satisfied by the function $\P$ \eqref{psieqn} as \eqref{delpsiinf}. Since the form of the boundary term is also the same in terms of the function $\P$, although the equation for the Kerr-schild scalar $\fks$, and therefore the metric function $f(r)$, becomes a polynomial equation that is considerably more complicated, the mass of the black hole $M$ is related to the integration constant $m$ in the same way given in \eqref{M}.

Instead of the relation \eqref{fromgrtom} for general relativity, we now have an analogous relation between the equation satisfied by the function $\psi$ \eqref{fksgen} denoted by $\text{EOM}(\psi)$ and the equation of motion of the scalar potential given in \eqref{phieqn2} [$\text{EOM}(\f)$] as
\begin{equation}\label{fromqtgtom}
\left[r\, \dv{r}-(D-4)\right] \text{EOM}(\psi) = \text{EOM}(\f).
\end{equation}
This suggests that the gravitational solution is mapped to the Coulomb solution provided that the following identification with the scalar potential $\f$ is made
\begin{equation}\label{ftofksgen}
\f = \psi = \sum_{n=1} ^{n_{\text{max}}} c_n \frac{\fks^n}{r^{2(n-1)}}, \quad \text{when } \f_\infty = 0.
\end{equation}
The relation between the electric charge $Q$ of the single copy and the black hole mass $M$ \eqref{QtoM} does not change since one has $\d \Psi = \d m$ for these theories although the function $\Psi$ is modified when quasi-topological Lagrangians are added to the action. Indeed, this is quite well-known for Lovelock theories and was first demonstrated in \cite{Crisostomo:2000bb}. By adding more quasi-topological Lagrangians to the action, one might have black hole solutions whose functional forms are quite different. However; we have shown that a gravitational solution is mapped to the Coulomb solution of Maxwell's theory with the identifications \eqref{ftofksgen} and \eqref{QtoM} provided that its reduced action is of the form \eqref{redGR}. For example, it was recently shown that one can obtain regular black hole solutions in a pure gravity theory by using infinitely many quasi-topological Lagrangians with carefully chosen coupling constants \cite{Bueno:2024dgm}, and as we have shown here, even those solutions are mapped to the Coulomb solution with these identifications. The effect of minimal matter coupling can be considered in the same way as explained for general relativity in the previous section. After this general proof, in the next section, we will provide an explicit realization in the simplest possible case beyond general relativity.

\section{5d Einstein-Gauss-Bonnet theory }\label{sec:EGB}
For the generalization of the Kerr-Schild double copy, we consider the Eintein-Gauss-Bonnet theory in $D=5$ with the action
\begin{equation}\label{actEGB}
I_{\text{EGB}} = \frac{1}{16 \pi G} \int \dd^5{x} \sqdet \left(R + \a \cL_2 \right), 
\end{equation}
where the second-order  Lovelock Lagrangian $\cL_2$, the Gauss-Bonnet invariant, reads from \eqref{Lm} as
\begin{equation}
\cL_2 = R_{\m\n\r\s} R^{\m\n\r\s}  - 4 R_{\m\n} R^{\m\n}+ R^2.
\end{equation}
Since the Gauss-Bonnet invariant identically vanishes in $D<4$, and is a topological invariant in $D=4$, this is the minimal case where the generalization is possible. The reduced action of the theory for the metric ansatz \eqref{BLwithN} is of the form \eqref{redGR}, where the $\P$ function, whose equation of motion is given by
\begin{equation}
\P = r^2 \left(\fks + 2 \a \frac{\fks^2}{r^2}\right) = m,
\end{equation}
matches with our general form \eqref{psigen} with $c_1=1$ and $c_2 = 2\a$. From this, one finds the explicit form of the Kerr-Schild scalar as
\begin{equation}
\fks = \frac{r^2 \pm \sqrt{r^4+8 \a m}}{4 \a},
\end{equation}
where the sign in front of the square root can be fixed to be minus by demanding a well-defined $\a \to 0$ limit. Our gravitational solution is mapped to the Coulomb solution with the identification
\begin{equation}\label{fksgb}
\f = \fks + 2 \a \frac{\fks^2}{r^2},
\end{equation}
and the conserved charges, the black hole mass $M$ and the electric charge $Q$, are related as in \eqref{QtoM}.

One can explicitly see this also in the standard formulation of the Kerr-Schild double copy. The covariant form of the field equations with a minimal matter coupling is as follows:
\begin{equation}
E_{\m\n} = G_{\m\n} + \a H_{\m\n} = 8 \pi G\, T_{\m\n},
\end{equation}
where the contribution from the Gauss-Bonnet term reads
\begin{equation}
H_{\m\n} = 2\left(R R_{\m\n} - 2 R_{\m\a} R_\n^{\ \a} - R^{\a\b} R_{\m\a\n\b} + R_{\m}^{\ \a\b\g} R_{\n\a\b\g}\right) - \frac{1}{2} g_{\m\n} \cL_2.
\end{equation}
Defining
\begin{equation}
\widetilde{E}_{\m\n} = E_{\m\n} - \frac{1}{3} g_{\m\n} E, \qquad E = E^\m_{\ \m},
\end{equation}
we can write
\begin{equation}\label{reversedegb}
\widetilde{E}^{\m}_{\ \n} = 8 \pi G\, \tT^{\m}_{\ \n},
\end{equation}
which is a generalization of the trace-reversed equations of general relativity \eqref{reversed}. Calculating the $\m0$-components in the Kerr-Schild coordinates by using (\ref{metKS}, \ref{back}, \ref{k}), one finds that the left-hand side  becomes
\begin{equation}
\widetilde{E}^{\m}_{\ 0} = -\frac{1}{2} \partial_{\nu} F^{\n\m},
\end{equation}
with the scalar potential given in \eqref{fksgb}. As a result, we obtain Maxwell equations with the source $J$ given in \eqref{source}.

It was reported in \cite{Anabalon:2009kq} that, while the field equations of a generic gravity theory derived from a Lagrangian containing quadratic curvature invariants are third-order in the Kerr-Schild scalar $\fks$, a simplification occurs for the Gauss-Bonnet combination and the field equations become quadratic. Here, thanks to our mini-superspace analysis, we are able to generalize the Kerr-Schild double copy construction and find that using the form of the field equations given in \eqref{reversedegb}, one can establish a map to Maxwell's theory. For quasi-topological gravities, considering how complicated the covariant form of the field equations are, repeating an analysis similar to that of \cite{Anabalon:2009kq} with the Kerr-Schild ansatz \eqref{metKS} seems practically impossible. However; as we have shown in the previous section, the double copy analysis, including sources, can be easily performed in the black hole mini-superspace. The only non-trivial part of the calculation is to determine the relevant coefficients $c_{n>1}$ in \eqref{psigen} in terms of the coupling constants in the action.
\section{Many-to-one nature of the generalized map and comparison with previous studies}\label{sec:many}

Thanks to the novel formulation of the classical double copy that can be implemented at the action level, a generalization to quasi-topological gravities was given. This might be a bit surprising at first sight because static, spherically symmetric black hole solutions of \emph{infinitely many} different gravity theories are mapped to the Coulomb solution of Maxwell's theory. When matter is added into the picture, they are still mapped to the same solution of Maxwell's theory modified by the source given in \eqref{source}. In this section, we show that this subtle behavior is observed only when one considers \emph{finite} higher-curvature corrections to the Einstein-Hilbert action. A more natural expectation would be to have a one-to-one map between solutions of different gravity and gauge theories. It turns out that such maps exist \emph{perturbatively}, which is not surprising considering the string theoretical origins of the double copy idea.

In the first attempt to address the nature of $\a^\prime$ corrections in the classical double copy  \cite{Pasarin:2020qoa}, a generalization of the Schwarzschild metric was constructed from an $\a^\prime$-corrected open string analogue of the Coulomb solution. It was approximated by a vacuum solution of Born-Infeld electrodynamics, which appears as the effective action of the abelian gauge field in the bosonic open string theory in the limit when the derivatives of the field strength tensor are small \cite{Fradkin:1985qd}. Assuming the Kerr-Schild ansatz \eqref{metKS} remain intact with $\fks = \f$ (no $\a^\prime$ or dilaton corrections), the authors constructed a double copy solution. They showed that although it seems regular, the corresponding curvature invariants still blow up at $r=0$. Therefore, the regularity of the gauge theory solution is not shared by its gravitational counterpart. Later, following the same procedure, regular double copy solutions were obtained from solutions of some non-linear electrodynamics theories with SO(2) electro-magnetic duality symmetry  \cite{Mkrtchyan:2022ulc}. In both works, the gravity theory admitting such double copy solutions was not given, which is natural since no guiding principle has been established yet. Furthermore, as emphasized in \cite{Pasarin:2020qoa}, the assumption regarding the Kerr-Schild ansatz might be too restrictive and such a gravitational theory might not exist at all. In order to make a comparison of this approach with ours, we will next consider a Born-Infeld-type extension of Maxwell's theory in 5d, which will also allow us to delve deeper into the many-to-one nature of the generalized map that we propose.

We consider the theory described by the action
\begin{equation}
I_{\text{BI}} = \frac{\b^2}{\O_3} \int \dd^5 x \sqrt{-\eta}\,  \left(1-\sqrt{1-\frac{F^2}{2 \b^2}}\right),
\end{equation}
which reduces to the action of Maxwell's theory in the limit $\b \to \infty$. Instead of the usual Born-Infeld action with a determinant, we can work with this simple form since it possesses the same static solution. The solution can be obtained by the mini-superspace method easily. Using the ansatz \eqref{ABL}, the reduced action can be obtained as
\begin{equation}\label{BIaction}
I_\text{BI} = \b^2 \int \dd{r} r^3 \left(1-\sqrt{1+\frac{{\f^\prime}^2}{2 \b^2}}\right).
\end{equation}
Note that we do not need to define a canonical momentum since we will not calculate the electric charge from scratch. This reduced action yields the following electric field and the scalar potential
\begin{equation}\label{solBI}
e = -\f^\prime = \frac{Q}{\sqrt{r^6+r^6_0}}, \qq \qq \f = -\b r \, _2F_1\left(\frac{1}{6},\frac{1}{2};\frac{7}{6};-\frac{r^6}{r^6_0}\right) \qq \qq r^6_0 = \frac{Q^2}{\b^2}.
\end{equation}
where $_2F_1$ is the standard hypergeometric function. One can easily verify that although the scalar potential and the metric function obtained from it ($f = 1-\f$) is regular at the origin, the curvature invariants show a singular behavior. As mentioned before, it is also not obvious how to construct a gravity theory admitting this black hole solution.

In order to understand some details, we can check what happens in a perturbative expansion in powers of $\nicefrac{1}{\b}$. The single copy properties in \eqref{solBI} have the following expansion near $\b \to \infty$
\begin{equation}\label{singpert}
e =  \frac{Q}{r^3}+\frac{Q^3}{2 \b^2 r^9}+\cO\left(\b^{-4}\right), \qq \qq \f = \frac{Q}{2 r^2}+\frac{Q^3}{16 \b^2 r^8}+\cO\left(\b^{-4}\right).
\end{equation}
The modification of Maxwell's theory admitting this solution can be found by expanding the action \eqref{BIaction} as
\begin{equation}
I_{\text{BI}} = \frac{1}{\O_3} \int \dd^5 x \sqrt{-\eta}\, \left[-\frac{1}{4} F^2 - \frac{1}{32 \b^2} (F^2)^2+\cO\left(\b^{-4}\right)\right].
\end{equation}

By using the minisuperspace approach, one can construct the gravitational theory whose solution is the double copy ($\fks = \f$) of the gauge theory solution \eqref{singpert} with relative ease. However, it is important to note that a perturbative expansion of the gravitational action in powers of curvature is not sensitive to the terms with the curvature scalar $R$ and the Ricci tensor $R_{\m\n}$. In order to see this, let us consider the following $\a^\prime$-corrected action
\begin{equation}
I = \int \dd^5 x \sqrt{-g} \left[R + \a^\prime \left( a_1  R_{\m\n\r\s} R^{\m\n\r\s} + a_2 R_{\m\n} R^{\m\n} + a_3 R^2\right) \right],
\end{equation}
where $a_1, a_2$ and $a_3$ are arbitrary constants. Under the field redefinition \cite{Brigante:2007nu,Kats:2007mq}
\begin{equation}
\delta g_{\m\n} = \a^\prime \left[a_2 R_{\m\n} - \frac{1}{3}(a_2+2 a_3) g_{\m\n}\right],
\end{equation}
one gets 
\begin{equation}
I = \int \dd^5 x \sqrt{-g} \left[R +  \a^\prime  R_{\m\n\r\s} R^{\m\n\r\s} + \cO\left({\a^\prime}^2\right) \right],
\end{equation}
where we have taken $a_1 = 1$. At each order in $\a^\prime$, one can use similar field redefinitions and end up with corrections containing the contractions of Riemann tensor $R_{\m\n\r\s}$ only and no terms with $R$ and $R_{\m\n}$. Therefore, the most general action for a gravitational theory up to the order ${\a^\prime}^2$ can be written as
\begin{equation}
I = \int \dd^5 x \sqrt{-g} \left[R + a_1 \a^\prime    R_{\m\n\r\s} R^{\m\n\r\s} \right. 
\left.+{\a^\prime}^2 \left( b_1 R^{\m\n\r\s} R^{\ \t \  \a}_{\m\ \r} R_{\n\a\s\t} + b_2 R^{\m\n\r\s} R^{\ \ \t \a}_{\m\n} R_{\r\s\t\a} \right) +   \cO\left({\a^\prime}^3\right) \right], 
\end{equation}
where $a_1, b_1, b_2$ are again arbitrary constants. Although the covariant form of the field equations are quite complicated, a perturbative solution can be easily constructed using the mini-space approach as
\begin{align}
	\fks =&\,\frac{m}{r^2} + \a^\prime \left[\frac{c_1}{r^2} - \frac{2 a_1 m^2}{r^6} \right]\nn\\ 
	&+{\a^\prime}^2 \left[\frac{c_2}{r^2} - \frac{4a_1c_1 m}{r^6}+ \frac{32 m^2 \left(16 a_1^2+3 (b_1-4 b_2)\right)}{r^{8}}\right. 
	\left. -\frac{2 m^3 \left(204 a_1^2+41 b_1-160 b_2\right)}{r^{10}} \right] \nn\\
	&+  \cO\left({\a^\prime}^3\right),
\end{align}
where $c_1$ and $c_2$ are integration constants. This can be matched with the perturbative gauge theory solution in \eqref{singpert} with the help of the following identifications
\begin{equation}
a_1=c_1=c_2=0, \qq \qq b_2 = \frac{41 b_1}{160}, \qq \qq m = \frac{Q}{2}, \qq \qq  \frac{1}{\b^2} =  \frac{4 b_1 {\a^\prime}^2}{5 Q}.
\end{equation}
Hence, the action for the ${\a^\prime}^2$-corrected gravitational theory should be taken as
\begin{equation}
I = \int \dd^5 x \sqrt{-g} \left[R +{\a^\prime}^2 \left( R^{\m\n\r\s} R^{\ \t \ \a}_{\m\ \r} R_{\n\a\s\t} + \frac{41}{160} R^{\m\n\r\s} R^{\ \  \t \a}_{\m\n} R_{\r\s\t\a} \right) +   \cO\left({\a^\prime}^3\right) \right], 
\end{equation}
where we have taken $b_1 = 1$ for simplicity. As we see, a one-to-one map is possible when perturbative corrections are considered.

Similarly, one can try to establish a one-to-one map between theories with higher-order corrections. At higher orders, the number of independent contractions of the Riemann tensor increases. Therefore, it seems reasonable to expect a match in all orders thanks to the increasing freedom in choosing the relative coefficients of the gravitational couplings. For the construction of the full gravity theory whose solution is the double copy of the solution in \eqref{solBI}, the following strategy might be useful: Considering a Born-Infeld-type gravity theory with some free parameters, one can hope to obtain a theory with less free parameters or a unique one after studying a few orders in the perturbative series. For example, in \cite{Alkac:2018whk}, a Born-Infeld-type gravity theory admitting a holographic c-theorem along the lines of \cite{Freedman:1999gp, Myers:2010xs, Myers:2010tj} was constructed starting from an appropriate ansatz. It is quite remarkable that the full theory can be obtained this way and it might even be a peculiar property of Born-Infeld-type theories among the non-linear theories.

It is now much easier to understand the many-to-one nature of our generalization because it arises when truncating the perturbative series in $\a^\prime$ and treating modifications of GR as finite corrections\footnote{This is essentially why the finite coupling constant corresponding to the Gauss-Bonnet Lagrangian in the action \eqref{actEGB} is denoted by $\alpha$, not $\a^\prime$.}. The need for adding terms with $R$ and $R_{\m\n}$ to the action to avoid ghosts in the low-energy limit of string theory is very well-known thanks to the seminal work of Zwiebach \cite{Zwiebach:1985uq}. Let us recapitulate the main argument here for completeness: Considering the fluctuations around the flat metric as $g_{\m\n} = \eta_{\mu \nu} + h_{\m\n}$ and using the harmonic gauge ($\partial_{\n} h^{\m\n} = \partial^{\m} h$ with $h = \eta_{\mu \nu} h^{\m\n}$), the contribution of the following combination of the quadratic-curvature terms in $D$-dimensions
\begin{equation}
I = \int \dd^D x \sqrt{-g} \left( a_1  R_{\m\n\r\s} R^{\m\n\r\s} + a_2 R_{\m\n} R^{\m\n} + a_3 R^2\right),
\end{equation}
is as follows
\begin{equation}
I = \frac{1}{4} \int \dd^D x \left[(4a_1 + a_2) h_{\m\n} \partial^2 h^{\m\n} + (a_3-a_1) h \partial^2 h \right] + \cO\left(h^3\right).
\end{equation}
To avoid ghosts, one needs to choose $a_2 = -4 a_1$ and $a_3 = a_1$, which leads to the Gauss-Bonnet Lagrangian that we studied in Section 5. This way, the theory that we consider describes a massless spin-2 graviton. The graviton propagator is not modified but the Gauss-Bonnet term affects vertices ($h^3$ with $n \leq 3$) as long as $D>4$. Quasi-topological gravities are a generalization of this idea to infinite-order in curvature, and indeed, \emph{any} gravitational effective action involving higher-curvature corrections is equivalent to a generalized quasi-topological gravity via metric field redefinitions \cite{Bueno:2019ltp,Bueno:2022res}. This way, one can truncate the effective field theory expansion at any order and consider a finite modification to GR which admits massless spin-2 excitations around a constant curvature vacua and static, spherically symmetric black hole solutions.

Thanks to the special form of the reduced action described in \eqref{redGR}, it becomes possible to map all solutions to the Coulomb solution of Maxwell's theory with the modified relation between the scalar potential $\f$ and the Kerr-Schild scalar $\fks$ given in \eqref{ftofksgen}. The Kerr-Schild scalar $\fks$ satisfies a first-order differential equation as given in \eqref{fksgen} for all these theories and it can be directly related to the equation \eqref{phieqn2} satisfied by the scalar potential $\f$ via \eqref{fromgrtom}. This would not be possible without the special combination of the higher-curvature terms (with $R$ and $R_{\m\n}$). Since there are no analogs of Lovelock or quasi-topological theories for Maxwell's theory, i.e., any modification with higher powers of $F^2$ yields higher-derivative field equations, it is singled out as the unique gauge theory whose solutions are the single copies of gravitational solutions that arise when certain finite corrections to GR are considered.
\section{Summary and outlook}\label{sec:sum}
In this paper, we gave a new formulation of the classical double copy for static, spherically symmetric black hole solutions, which are mapped to purely electric solutions of Maxwell's theory. By obtaining one-dimensional mini-superspace models for such solutions, we proved its full equivalence to the Kerr-Schild double copy for solutions of general relativity minimally coupled to matter. This formulation has the advantage that it allows for an extension of the classical double copy to quasi-topological gravities, which are the higher-curvature extensions of general relativity in $D>4$ that share important physical properties, as explained in section \ref{sec:QTG}. In this generalization, the scalar potential $\f$ in Maxwell's theory becomes a polynomial of the Kerr-Schild scalar $\fks$ corresponding to the gravitational solution as described in \eqref{ftofksgen}. By calculating the Regge-Teitelboim surface charges, we proved that the relation between the black hole mass $M$ and the electric charge $Q$ corresponding to the Coulomb part of the gauge theory solution is universal [see eq. \eqref{QtoM}]. Finally, a detailed discussion of the many-to-one nature of this generalized map was given in Section \ref{sec:many}.

We believe that it should be possible to generalize our construction to solutions that can be written in the Kerr-Schild coordinates with a constant-curvature \cite{CarrilloGonzalez:2017iyj} and a generic curved background \cite{Alkac:2021bav}. Especially in the latter case, when the Kerr-Schild ansatz with such backgrounds is inserted into the covariant form of the trace-reversed equations \eqref{reversed}, the resulting equations are considerably involved. Since our mini-superspace approach directly imposes the symmetries of the solution under consideration in the reduced action, it is expected to be more practical in such cases. Additionally, by employing the Euclidean approach to thermodynamics \cite{Gibbons:1976ue}, the reduced action can be used to deduce the thermodynamical properties of black hole solutions (see, for example, \cite{Crisostomo:2000bb}). Therefore, one might look for an interpretation of these properties in Maxwell's theory easily in the mini-superspace approach. We plan to address these issues in our future works.


{\bf \sffamily Acknowledgments} G. A. and M. K. G. are supported by T\"{U}B\.{I}TAK Grant No 124F058.

\end{document}